# Two Types of 3D Quantum Hall Effects in Multilayer WTe$_2$

Xurui Zhang[1], Yeonghun Lee[2,3], Vivek Kakani[1], Kun Yang[4], Kyeongjae Cho[2], Xiaoyan Shi[1*]

[1]*Department of Physics, The University of Texas at Dallas, Richardson, TX, USA*

[2]*Department of Materials Science and Engineering, The University of Texas at Dallas, Richardson, TX, USA*

[3] *Department of Electronics Engineering, Incheon National University, Incheon, Republic of Korea*

[4]*Department of Physics and National High Magnetic Field Laboratory, Florida State University, Tallahassee, FL, USA.*

## Abstract

Interplay between the topological surface states and bulk states gives rise to diverse exotic transport phenomena in topological materials. The recently proposed Weyl orbit in topological semimetals in the presence of magnetic field is a remarkable example. This novel closed magnetic orbit consists of Fermi arcs on two spatially separated sample surfaces which are connected by the bulk chiral zero mode, which can contribute to transport. Here we report Shubnikov-de Haas (SdH) oscillation and its evolution into quantum Hall effect (QHE) in multilayered type-II Weyl semimetal WTe$_2$. We observe both the three-dimensional (3D) QHE from bulk states by parallelly stacking of confined two-dimensional layers in the low magnetic field region and the 3D QHE in the quantized surface transport due to Weyl orbits in the high magnetic field region. Our study of the two types of novel QHEs controlled by magnetic field and our demonstration of the crossover between quantized bulk and surface transport provide an essential platform for the future quantized transport studies in topological semimetals.

## Introduction

A two-dimensional electron gas (2DEG) in a perpendicular magnetic field exhibits the quantum Hall effect (QHE)[1], in which the Hall conductivity $\sigma_{xy}$ will be quantized at $\nu e^2/h$, where the integer $\nu$ is the filling factor, indicating the number of filled Landau levels (LLs)[2,3], and $e$ and $h$ are the electron charge and the Planck constant, respectively. The quantized $\sigma_{xy}$ always accompanies simultaneously the vanishing longitudinal resistance $R_{xx}$. This is attributed to the dissipationless transport along the chiral edges of the sample. In contrast, the bulk of the sample is insulating. Such quantum Hall systems are very special types of topologically non-trivial insulating states discovered later[4,5], which are members of a very large family of topological quantum states[6]. The origin of quantization of $\sigma_{xy}$ lies in the non-trivial topology, with ν being a topological invariant which is independent of material type and geometry[2,3]. In the past few years, predicting and realizing the QHE in topological materials have attracted intense interest due to both the rich physics behind it and potential applications in quantum devices. It has been reported that, in the prototype 3D topological insulator (TI) $Bi_2Se_3$, the QHE exhibits strong thickness-dependence, which is called as quasi-2D QHE[7]. In addition, theories suggest that a half-quantized surface Hall conductance representing the topologically nontrivial nature can be induced by breaking the time reversal symmetry (TRS) in TIs[5,8]. Indeed, such a half-integer QHE has been observed in an intrinsic three-dimensional (3D) TI material $BiSbTeSe_2$[9]. Furthermore, recent theoretical studies predict formation of an exotic orbit, the so-called Weyl orbit, can support the QHE in the 3D gapless topological materials, Dirac and Weyl semimetals (DSM/WSM)[10–13]. In such materials, there exist topologically protected surface states that form open Fermi arcs, connecting pairs of bulk Weyl nodes with opposite chiralities[14,15]. In the presence of magnetic field, the two Fermi arcs at opposite sample surfaces are connected by bulk chiral modes to form closed Weyl orbits[10–13,16–18], which can support the QHE in bulk (3D) crystals. There has been some experimental verification of the QHE in 3D DSM $Cd_3As_2$[19].

Tungsten ditelluride ($WTe_2$), a layered transition metal dichalcogenide (TMD) material, has attracted a great deal of interests due to its unique topological properties. In the bulk form, $WTe_2$ has been predicted[20] and observed[21] to be a type-II Weyl semimetal. Interestingly, the monolayer form of $WTe_2$ is a quantum spin Hall insulator (QSHI) at low carrier density ($n$)[22,23]. However, QHE, as the first example of a topological quantum state, has never been observed in this material due to the limited carrier mobilities[24–27]. In this paper, we report the QHE in a multilayer $WTe_2$ film with low $n$. The bulk states contribute to the 3D QHE, in which the thickness-dependent quantization comes from parallel 2DEG conduction channels due to the quantum confinement along the normal direction of sample surface. With the magnetic field increasing, the bulk chiral LL (0$^{th}$ LL) mixes with the surface states that form the Fermi arc, which build up the Weyl orbits. We find that the quantization of these orbits gives rise to further quantization of Hall conductance.

## Experiment

The multilayer WTe$_2$ flake was obtained by mechanical exfoliation of bulk WTe$_2$ crystals purchased from HQ graphene company. Two pieces of hexagonal Boron Nitride (hBN) thin flakes were used to encapsulate the WTe$_2$ flake and transferred onto a silicon substrate with 285 nm SiO$_2$ coating on surface by dry transfer technique[28]. The hBN flakes are necessary here to prevent the WTe$_2$ flake from being oxidized in the air[29,30]. In addition, they provide a cleaner interface and improve carrier mobility. A standard fabrication process was applied to make electrical contacts to the stacking structure to make a Hall bar pattern. Specifically, contacts were defined by electron beam lithography with PMMA 950K A4 photoresist and developed in MIBK/IPA solution. The top hBN layer in the contact region was removed by reactive ion etching using CF4/Ar chemistry. Immediately thereafter, Pd/Au (10 nm/50 nm) layers were deposited in an e-beam deposition chamber with high vacuum. Lift-off process of excess metal was done by immersing the sample in acetone at room temperature for 2 hours. The device was then wire bonded and attached to a DIP socket for transport measurements. An optical image of the WTe$_2$ device is shown in Fig. 1a. Transport measurements down to $20\ m\text{K}$ were carried out in an Oxford dilution refrigerator. Sample thickness (24.2 nm) was determined after transport measurements by using an AFM (Fig. 1b).

To calculate band structures of WTe$_2$, we utilized the density functional theory (DFT) implemented in Vienna Ab initio Simulation Package (VASP)[31,32]. The exchange-correlation energy functional was given by the Perdew-Burke-Ernzerhof (PBE) functional[33] in the generalized gradient approximation (GGA). The pseudopotential was given by the projector-augmented wave (PAW) method[34,35]. The energy cutoff for the plane-wave basis set was 250 eV, and a 7 × 3 × 1 Γ-centered $k$-point grid was adopted for the Brillouin zone sampling. Spin-orbit coupling was considered. The zero damping DFT-D3 method of Grimme[36] was employed to describe van der Waals interactions. The cell geometry was optimized until the maximum atomic force is smaller than 0.01 eV/Å.

The magnetotransport simulation of QHE of a five-layer system was performed by using Kwant, a software package for quantum transport[37]. The multilayer system was described by the tight-binding model along with the 20 × 20 × 5 cubic lattice structure with a single $s$-orbital without spin degeneracy, where the hopping Hamiltonian matrix elements between nearest neighbors were set to -1 Ry, and the interlayer coupling is assumed to be small (-0.01 Ry) to mimic a 3D bulk. The system subjected to a perpendicular magnetic field was modeled by Peierls substitution[38,39].

## Results and discussions

We first measured the temperature ($T$) dependence of the longitudinal resistance $R_{xx}(T)$ at zero magnetic field. This sample resistance $R$ increases as $T$ decreases at $T > 30$ K (*i.e.*, insulating behaviour), while $R$ decreases as $T$ increases at $T < 30$ K (metallic behaviour). As shown in Fig. 1c, an anomalous peak emerges at around 30 K and can be ascribed to the temperature-dependent Fermi energy shift of the electronic band structure. Such feature, as a hallmark in $ZrTe_5$ which is a 3D quantum Hall material, has been proved to be caused by changes in carrier types, which is intrinsically due to the temperature-induced Fermi energy shift of the electronic band structure.[40,41]. It is reasonable that this can happen in $WTe_2$, because it has been confirmed that the Lifshitz transition occurs in $WTe_2$ in a large temperature range[42]. If the carrier density is low enough, insulating behavior dominated by a certain type of carrier could also happen. To verify this, elaborate measurements of angle-dependent magnetoresistance (MR) and Hall effect provide further insight into the band-topological properties, as shown in Fig. 1d and 1e, respectively.

Magneto-transport with rotational magnetic field is conducted to detect the MR in $WTe_2$. In the Hall bar sample, the current ($I$) was applied perpendicular to the magnetic field ($H$), which is applied along the stacking direction of the WTe$_2$ layers. A schematic illustration of the sample and external $H$ field is shown in the inset of Fig. 1e. The most conspicuous feature is that, when the magnetic field $\mu_0 H > 1$ T is applied perpendicular to the current, clear Shubnikov-de Haas (SdH) oscillations can be identified in MR and several plateaus can be clearly observed in Hall measurements, indicating a QHE in such a high-quality sample with both a low carrier density ($n$) and a high carrier mobility ($\mu$). By fitting the low-field linear regime in Hall measurement and combined with the normal-state longitudinal resistance at zero angle, we extract the electron-type carrier density to be $3.2 \times 10^{18}$ cm$^{-3}$ and mobility to be $32400$ cm$^{-2}$/Vs. The carrier density we obtained is at least an order of magnitude lower than those in conventional WTe$_2$ flakes and the mobility is an order of magnitude larger[25–27]. As the magnetic field is tilted away from perpendicular position with an angle ($\beta$), the MR damps with the law of cosines, likely suggesting a 2D nature and a highly anisotropic Fermi surface with the cigar or ellipsoid shape. This is reasonable for a layered material, and further transport study along the stacking direction can be helpful to determine the shape of Fermi surface. We also found that the absolute resistance values of those plateaus and oscillations are independent of the tilted angle which is a hallmark of conventional QHE. Details about the QHE will be discussed in the following parts.

We extracted the oscillations in $R_{xx}$ curves at different angles by removing the large positive MR background, which is an intrinsic characteristic in Weyl semimetal WTe$_2$[43]. The extracted oscillation $\Delta R_{xx}$ is plotted versus $1/\mu_0 H \cos\beta$ in Fig. 2a. An anomalous feature becomes evident, namely the periodicity clearly changes at large fields, which can be observed in Fig. 2a. From the Fourier transformation of the SdH oscillations, as shown in Fig. 2b, it is clear that there exist higher oscillation frequencies in the large field region, in addition to the small primary frequency which dominates the low field region as the only frequency at around 7.5 T. We first focus on the low field region. A Landau fan diagram is plotted by following maxima (associated with integers) and minima (associated with half-integers) in the SdH oscillations, as shown in Fig. 2c. The oscillation frequency $B_F$ and Berry phase $\phi_B$ can be extracted according to the Lifshitz–Onsager quantization rule[44]: $\frac{B_F}{B} = N + \gamma = N + \frac{1}{2} - \frac{\phi_B}{2\pi} - \delta$, where $B = \mu_0 H$, $N$ is the Landau level (LL) index,

$\gamma$ is the intercept in the fan diagram, based on the expression of the oscillating term, and $\delta$ is an additional phase shift determined by the dimensionality, taking the value $\delta = 0$ (or $\pm 1/8$) for 2D (or 3D) case[44]. In addition, for Dirac fermions, a value of $|\gamma|$ between 0 and 1/8 implies a non-trivial $\pi$ Berry phase, whereas a value of around 1/2 represents a trivial Berry phase[41]. The intercepts at different angles are shown in the inset of Fig. 2c. When the magnetic field is perpendicular ($\beta = 0°$), the Landau fan diagram yields an intercept $\gamma$ around $1/8$, exhibiting a non-trivial Berry phase for the detected Fermi surface (Fig. 2d). However, as the magnetic field is further rotated to almost parallel to the current ($\beta = 87°$ and $90°$), the Berry phase begins to deviate from the non-trivial and finally turns to be trivial ($\gamma \sim 1/2$). This suggests that, when the magnetic field is applied along z direction, the Berry phase lies in the region of $\pi \pm \pi/8$, which makes the Landau fan plot exhibit an intercept between $\pm 1/8$. The Berry phase for cyclotron orbits corresponding to magnetic field along current direction is small and corresponds to an intercept close to 1/2. Such behavior agrees with a weak topological insulator[41], indicating the topological state in WTe$_2$. Moreover, the exact intercepts at $\pm 1/8$ suggest the 3D nature of the Fermi surface, which indicates the unconventional QHE in our sample. We have also obtained the angular dependence of $B_F$ as illustrated in Fig. 2e where a better agreement with 3D case ($B_F^{3D} = B_{F,z}B_{F,//}/\sqrt{(B_{F,z}sin\beta)^2 + (B_{F,//}cos\beta)^2}$, $B_{F,i}$ is the frequency of the SdH oscillation in different direction) is reached, rather than a simple $1/cos\beta$ relationship in 2D case, confirming the 3D nature of the Fermi surface[41].

Now we focus on the QHE in the multilayer WTe$_2$. Fig. 3 summarizes the magnetotransport for a multilayer WTe$_2$ sample. In Fig. 3a, SdH oscillations and corresponding plateaus can be resolved in both longitudinal resistance $R_{xx}$ and Hall resistance $R_{xy}$. We mark the integer LL index $N$ where the Hall plateau is accompanied by an obvious dip in the longitudinal $R_{xx}$, which provides clear evidence for the QHE. Furthermore, we notice that the value of the quantized $R_{xy}$ on the plateau is much smaller than the von Klitzing constant $h/e^2$ by a factor of 42, which applies for all the observed plateaus. Such behavior is common in 3D QHE[7,40,41,45,46] and can be simply explained for layered materials as follows. The sub-bands formed from the quantum confinement along the layer stacking direction are responsible for this large factor. The value of the factor depends on the number of the sub-bands that are occupied, which is determined by the confinement thickness, and also the degeneracy factor[47]. Given the weak band dispersion along the stacking direction in such van der Waals material, one can expect the number of subbands contributing the QHE to be the same as the number of layers. Alternatively, one can view the observed QHE as the result of many parallel 2D conduction channels, each acting as a 2D electron system, making up the stacking 3D system, which is similar to the case in bulk $Bi_2Se_3$ [7]. We emphasize this does not mean the interlayer hopping is not important in our system; to the exact opposite it is crucial to the topologically non-trivial 3D band structure and the presence of Weyl nodes. To verify our interpretation, we plot normalized inverse Hall resistance $R_{xy_0}/R_{xy}$ versus $B_F/B$ at 20 mK. $1/R_{xy_0}$ is the step size between the consecutive plateaus determined from the plateau values at $\frac{B_F}{B} = 1, 2, 3$. The inverse of $R_{xy}$ also exhibits clear plateaus at regular intervals of $B_F/B$, the positions of which nicely correspond to deep minima in $R_{xx}$. Given the conventional view of QHE, this quantization of $R_{xy_0}/R_{xy}$ leads to the normalized filling factor or LL index $N$. This is consistent with the plateaus occurring at integer multiples of $B_F/B$, where a small shift can be attributed to the Berry phase. Such a normalized filling factor is known to stem from the nontrivial $\pi$ Berry phase of Dirac fermions, which in two dimensions leads to the Hall resistance quantized as follows[45]

$$\frac{1}{R_{xy}} = \pm s(N + \gamma)\frac{e^2}{h}$$

where $s$ is the degeneracy factor. This gives the step size between the successive $1/R_{xy}$ plateaus as $1/R_{xy_0} = sZ(\frac{e^2}{h})$, $Z$ is the total number of the stacking layers along the $c$ axis. We have performed a DFT band calculation in zero field to investigate the degeneracy of $WTe_2$ under applied strain, though the calculation cannot provide further insight into the quantum Hall data reported here. From the DFT band calculation, as shown in Fig. 3c, the degeneracy factor $s$ has been reduced from 4 in the unstrained band to 2 in the strained one, and $s$ can be further reduced to $s = 1$ if the Kramer's degeneracy has been lifted. In Weyl semimetal WTe$_2$, the spatial-inversion symmetry has already been broken. The application of external field will lead to the naturally lifted Kramer's degeneracy[48]. Then we get $Z = 42$ ($sZ = 42$), indicating the 3D system is made up of 42 layers of 2D electron gas in parallel. We extract the 2D carrier density for each parallel 2D conduction channel from the SdH oscillations as $n_{2D}^{SdH} = \frac{e\,s\,B_F}{h} = 1.91 \times 10^{11}\ cm^{-2}$. Recall that the bulk carrier density extracted from Hall measurement to be $n_{3D}^{Hall} = 3.2 \times 10^{18}\ cm^{-3}$. The ratio between them allows us to predict the effective thickness ($t_{2D}$) per 2D conduction channel as $t_{2D} = \frac{n_{2D}^{SdH}}{n_{3D}^{Hall}} = 0.6\ nm$, which is close to but slightly smaller than the theoretical thickness of 1 layer of WTe$_2$ ($0.7\ nm$). Considering the 3D system is made up of 42 layers of conduction channel, we could get the direct measurement of each layer thickness to be $t_{2D} = t/42 = 0.58\ nm$ from the AFM result ($t$ = 24.2 nm), which is surprisingly consistent with the aforementioned prediction and the Hall plateau resistance. It is therefore proved that the 3D QHE in our sample consists of 42 parallel 2D conduction channels, each of which is a physical layer.

Now we go back to the QHE at high field region (> 9 T). As indicated in Fig. 2b, the single oscillation frequency characterized from the low field region corresponds to the 3D bulk Fermi surface that has been discussed above. However, much larger oscillation frequencies, corresponding to larger Fermi surface area, is superimposed on that of the bulk state from high field region. The integer QHE originated from the bulk state clearly stretch down to the quantum limit with LL index $N = 1$ at around 7 T (Fig. 3a). That means the bulk carriers are already confined to the lowest LL and can no longer produce any quantum oscillations. However, as shown in Fig. 4c, clear oscillations can be observed in the second order differential $\frac{d^2R_{xx}}{dH^2}$ and several quantized plateaus can be seen corresponding to the oscillation valleys, indicating that the additional QHE observed in the high-field MR is of 2D nature developed from the surface states. Although as a Weyl semimetal, the surface states of WTe$_2$ exist in the form of open Fermi arcs, which cannot support closed magnetic orbit individually, the bulk chiral zero mode can bridge the Fermi arcs on the opposite surfaces to form closed Weyl orbits[10,13]. The schematic plot of the corresponding energy-band structure and the transport behavior of Weyl orbit in a slab with thickness $t$ are shown in Fig. 4a and 4b, respectively. The experimental observation of Weyl orbit has been verified in a DSM Cd$_3$As$_2$[49–52]. Fig. 4d and 4e show temperature dependence of the extracted oscillatory components in $R_{xx}$ and the effective mass $m^*$ extracted from the fitting by Lifshitz–Kosevich (LK) formula [44] as a function of $\frac{1}{\mu_0 H}$. In the low field region, $m^*$ is generally below $0.1 m_e$ ($m_e$ is the mass of free electron). Such a small effective mass agrees well with previously reported Dirac or Weyl semimetals[53]. By contrast, in the high field region, $m^*$ becomes several times larger ($0.4 \sim 0.5 m_e$) for the higher-frequency oscillations. According to the LK formula, larger effective mass from the surface state results in a quicker decay of the oscillation amplitude. Such behavior of the surface state has been observed in previous transport studies of Weyl orbit[52,54]. In our sample, the estimated mean free path, $l = \frac{\mu m^* v_F}{e} \approx 700 \ nm$, fulfills the requirement for observing the Weyl orbit ($l > t$)[10]. In addition, we estimated the oscillation frequency of the Weyl orbit, $F_s$, in the sample based on the theory[10]. $F_s = E_F k_0 / (e \pi \bar{v}_F)$, where Fermi energy $E_F = \hbar \bar{v}_F \bar{k}_F$ and Fermi arc separation $k_0 \sim 0.32 \pm 0.02 \ nm^{-1}$ (Ref. [55]). Based on the charge density extracted from the Hall experiment and SdH oscillations, we have the average Fermi wave vector $\bar{k}_F = 1.0 \ nm^{-1}$ (assuming a circular 2D Fermi surface) and the calculated quantum oscillation caused by Weyl orbit is 70 T, which agrees well with our experimental observation of ~75 T (Fig. 2b). It is worth noting that this experimental observation also agrees with reported 78 T in a WTe$_2$ ribbon[21]. While the surface QHE based on Weyl orbit occurs at higher fields, the bulk states reach quantum limit and support the Weyl orbit as an inter-connection. This is consistent with recent theoretical work[56,57] suggesting that the Weyl nodes of a Weyl semimetal are robust and re-emerge in the presence strong magnetic field that reorganizes the original bands into Chern bands responsible for the 3D QHE before quantum limit is reached. According to the theory of Weyl orbit, the Hall conductance $\sigma_{xy}$ depending on both the surface states and the bulk chiral Landau bands can be written as $\sigma_{xy} = \frac{e^2}{h}(n_0 + n)$, where $n$ is an integer and the additional $n_0$, which originates from the bulk chiral Landau bands, takes the form of $n_0 = \lfloor \frac{k_w t}{\pi} \tan\beta \rfloor$, where $k_w$ is the position of Weyl nodes along $B$ direction in momentum space[13]. In perpendicular $B$, $tan\beta = 0$ and $n_0 = 0$, the Hall conductance degenerate into the general form $\sigma_{xy} = \frac{ne^2}{h}$. The step size between the consecutive plateaus becomes one conductance quantum, as shown in Fig. 4c. Please note, there is no degeneracy factor *Zs* in this plot. It is also worth mentioning that the much smaller bulk effective mass can very easily induce the quantum confinement and sub-bands formation along the $c$ axis, thus leading to the 3D QHE in bulk state[47]. That's the reason why the bulk transport is confined in each layer vertically. We believe the Weyl orbits in type-II WSM (with its unique electronic structure) is likely to be much more complicated than those in conventional DSM and type-I WSM, especially when considering the relative field orientations and the lattice anisotropy. Therefore, more theoretical and experimental works are desired to further clarify the detailed nature of the underlying mechanism of the surface orbits in WTe$_2$.

In conclusion, we have observed two different types of 3D QHEs co-existing and the crossover from one to the other with increasing field in multilayer WTe$_2$ flakes by fabricating high quality samples with high carrier mobility. Through detailed analysis of the SdH oscillation and Hall quantization, we divide the observed QHEs into a 3D bulk state and a surface state based on Weyl orbit according to the strength of magnetic field. In low field region, the quantum confinement is responsible for the gap opening and formation of sub-bands of bulk states. The magnetotransport is confined in each physical layer and results in the multiple parallel 2D conduction channels. In high field region, under the support of bulk chiral zeroth mode, the Weyl orbits, consisting of the Fermi arcs on opposite sample surfaces and the bulk chiral Landau level, are formed to conduct the surface QHE. Unlike the DSM and type-I WSM where experiments on QHE have been reported [10–13,16,19], further investigations are desired for more details of the underlying physics in WTe$_2$, a typical type-II WSM, such as Fermi surface topology, role of electron-electron interactions, non-local transport and scanning probe microscopy for the surface conduction, electric gating for the gap structure, designing hetero-interfaces for independently surface control. In this respect, our investigation of the 3D bulk and surface QHE in a type-II WSM provides an essential platform for the novel quantized transport studies in the future.

**Acknowledgments**

This work of X.Z., Y.L., V.K. and X.S. was supported by UT Dallas SPIRe fund (No. 2108630). The work of K.Y. was supported by the National Science Foundation (Grant No. DMR-1932796), and performed at the National High Magnetic Field Laboratory, which is supported by National Science Foundation Cooperative Agreement No. DMR-1644779, and the State of Florida.

**Data availability**

The data that support the findings of this study are available from the corresponding author upon reasonable request from the authors.

* Email: xshi@utdallas.edu

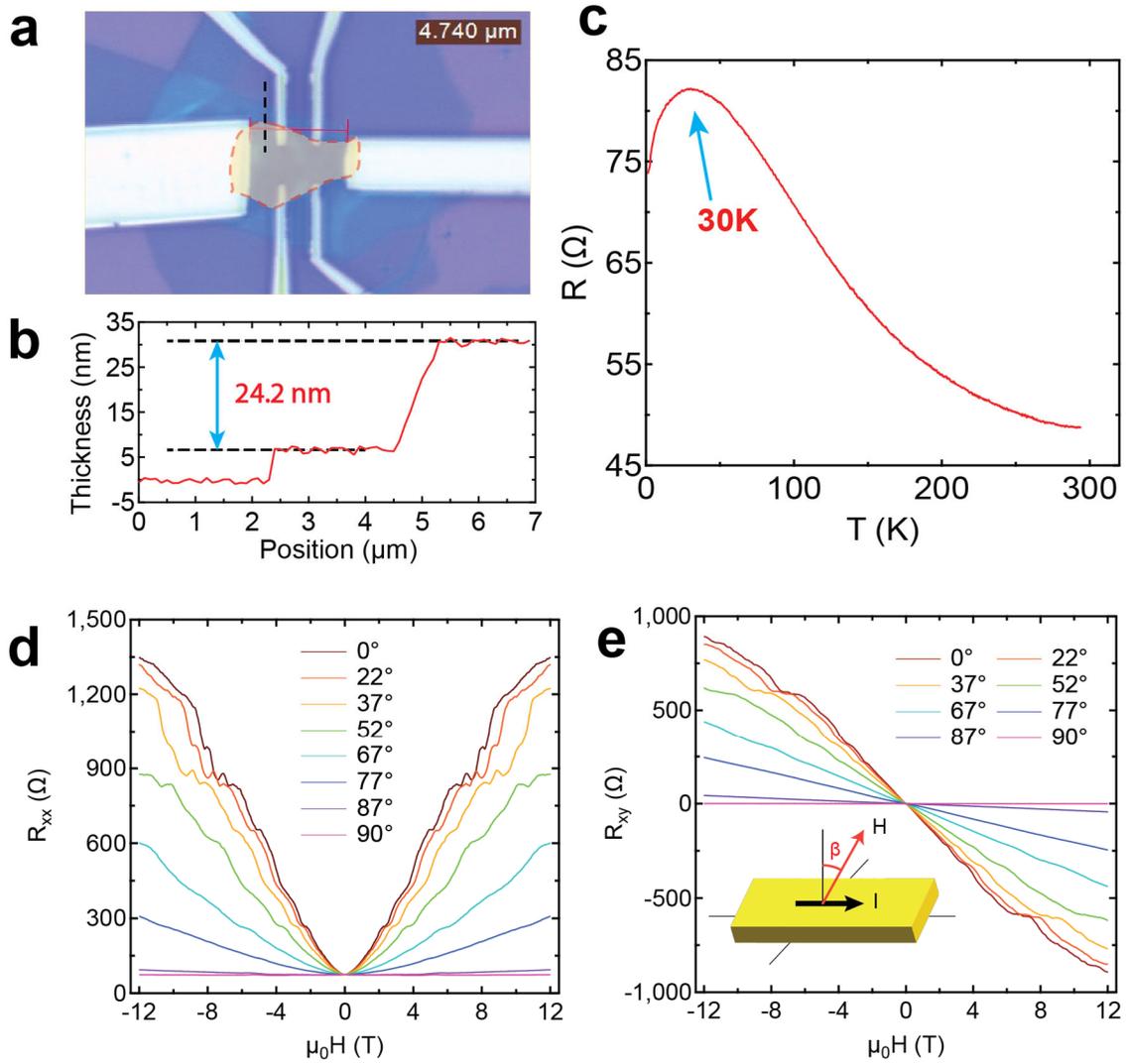

**Fig. 1 a,** An optical image of the WTe$_2$ device. The red scale bar gives the length of channel which is 4.74 µm. The black dashed line guides the AFM measurement. WTe2 flake is highlighted as the orange shaded area. **b,** Sample thickness is 24.2 nm determined by AFM measurement. The first step measures the thickness of the bottom hBN. **c,** Temperature dependence of the longitudinal resistance $R_{xx}(T)$ at zero magnetic field. The anomalous resistance peak occurs around $T_p \approx 30$ K. **d,e,** Angular dependence of MR and Hall resistance at 20 mK, respectively. Symmetrized data are shown in (d) and (e). The inset in **e** shows a schematic illustration of the sample and external magnetic field. β is the angle labeled in **d** and **e**.

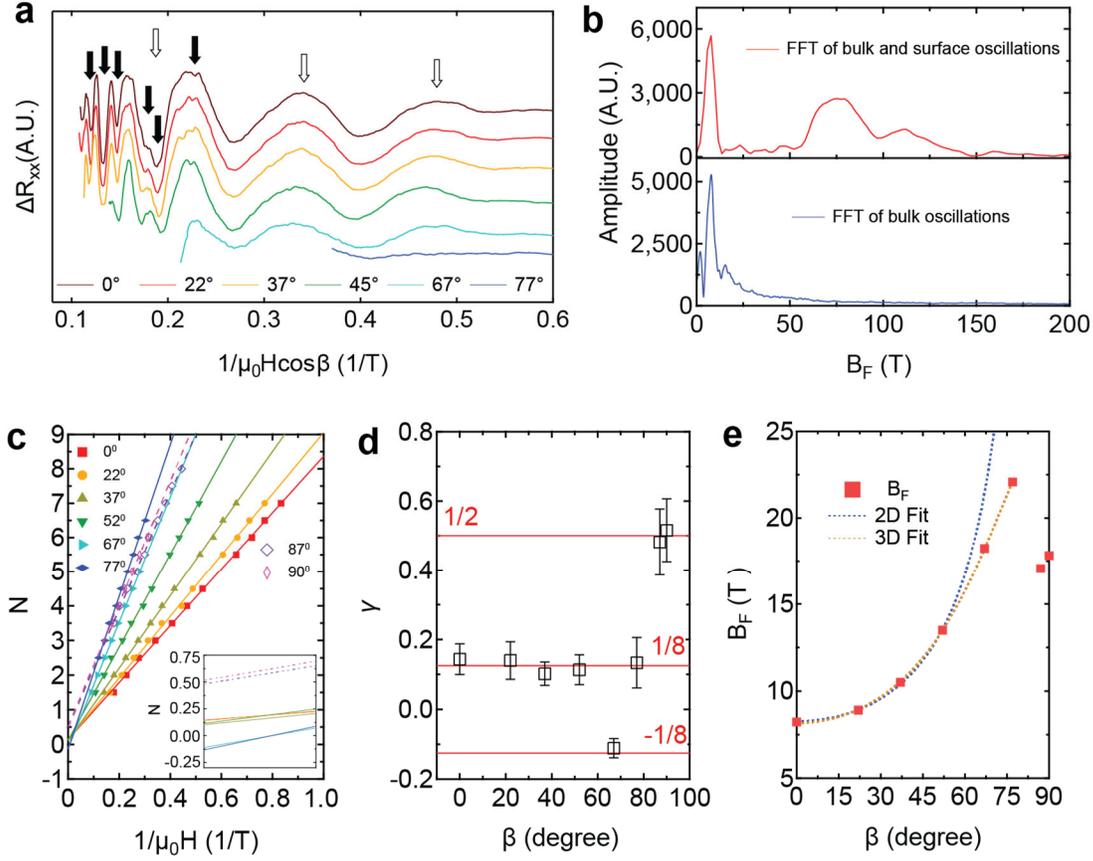

**Fig. 2 a,** Angle dependence of the extracted oscillations $\Delta R_{xx}$ versus $1/\mu_0 H cos\beta$. Curves are shifted vertically for clarity. Hollow and filled black arrows denote the bulk and surface oscillation components, respectively. **b,** Fourier transformation of the SdH oscillations in whole field region (upper) and only low field region (lower), respectively. **c,** Landau fan diagram of the SdH oscillations at different angle $\beta$, and its extrapolation to evaluate the intercepts. Inset is the zoom in plot near the origin of the main plot (0,0). A sidenote here, to better determine the intercept values from the fan diagram, smaller angle step is needed to compensate the drastic decreasing of effective field ($\sim \mu_0 H cos(\beta)$) near 90° angle. **d,** The angular-dependent intercept of Landau fan diagram in **c**. The error bars were generated from the linear fitting process in the Landau fan diagrams. **e,** Angle dependence of oscillation frequency $B_F$ as a function of $\beta$. The blue and golden dashed curves represent the fittings in 2D and 3D cases, respectively.

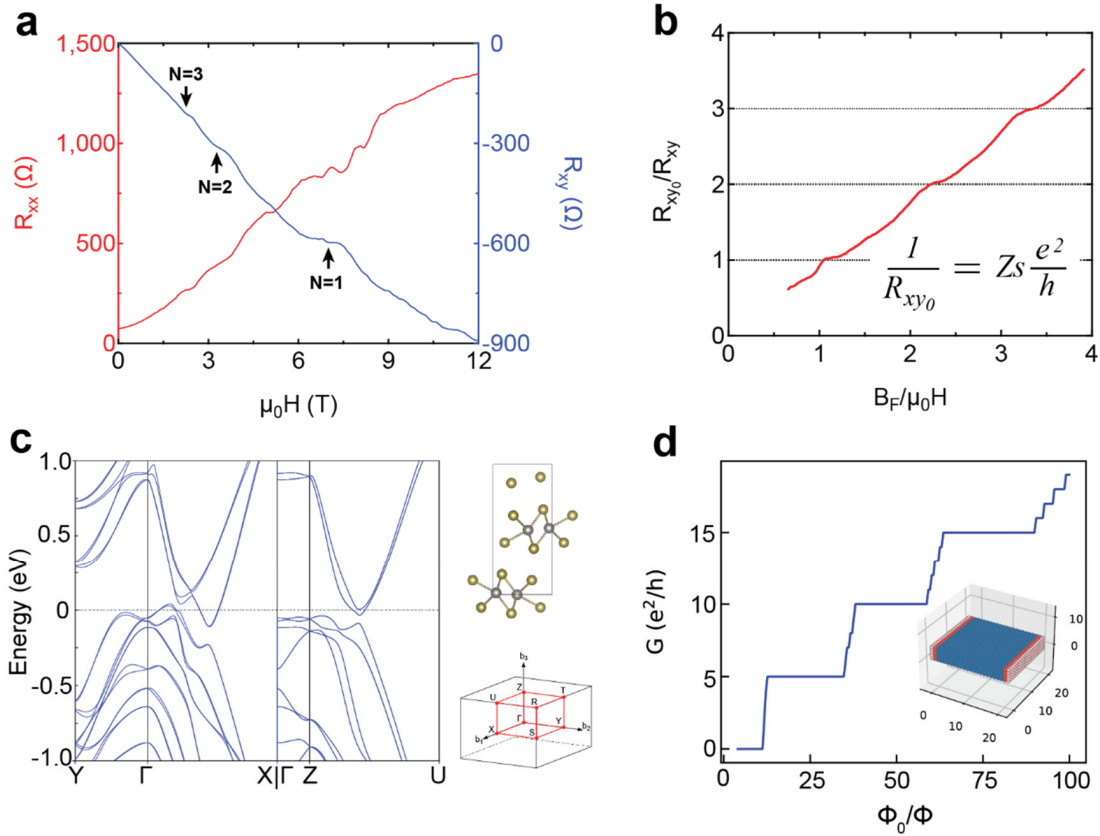

**Fig. 3** Quantum Hall effect observed in WTe$_2$. **a,** Magnetotransport for the WTe$_2$ multilayer measured at 20 mK. The red arrows mark the quantized plateaus in $R_{xy}$ and oscillation valleys in $R_{xx}$. LL index ($N$) are assigned. **b,** Normalized inverse Hall resistance $R_{xy_0}/R_{xy}$ versus $B_F/B$ measured at 20 mK. $1/R_{xy_0}$ is the step size between the consecutive plateaus. Degeneracy factor $Zs$ = 42. **c,** DFT band calculation of bulk WTe$_2$ in zero field. 1% tensile strain along X direction is applied to get the minimum degeneracy factor and reflect the possible strain in the real device. The upper inset shows one unit cell of WTe$_2$, which contains two physical layers, where the X direction is perpendicular of the plane of the paper. The lower inset shows the orthorhombic Brillouin zone in momentum space. **d,** Hall quantization shown in the magnetotransport simulation of a 5-layer flake. $\Phi_0$ is the flux quantum $h/(2e)$, where even in the presence of interlayer coupling, the first few plateaus remain intact. The inset shows the artificial sample in real space.

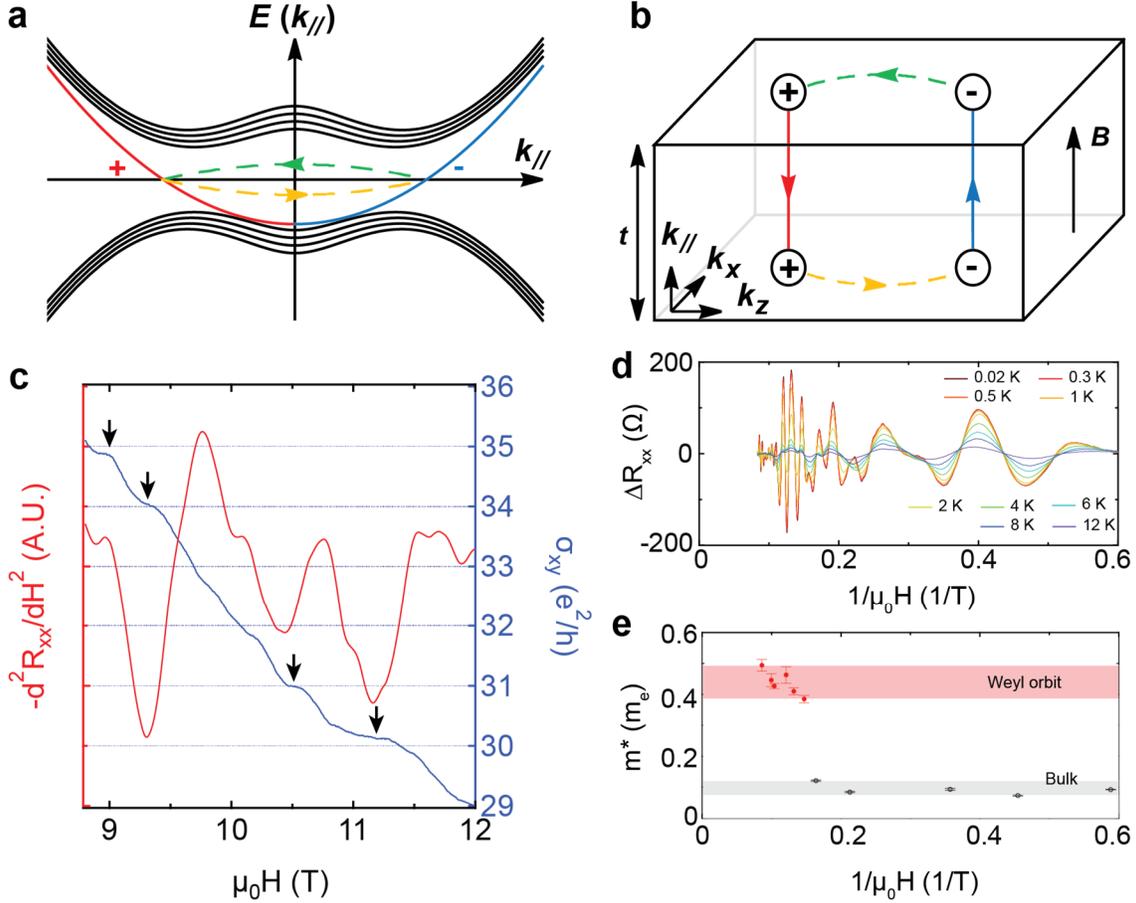

**Fig. 4** Quantum Hall effect from Weyl orbits. **a,** The schematic plot of the energy-band structure under a magnetic field $B$. $k_{//}$ is the wave vector parallel to $B$. The red and blue solid curves represent the chiral bulk Landau bands. The yellow (green) dashed curve represents the Fermi arc on the bottom (top) surface. **b,** The schematic plot of Weyl orbits in a sample slab with thickness $t$. The two Fermi-arcs on the top and bottom surfaces are interconnected through the bulk chiral zeroth mode. The colors of the curves correspond to the colors in **a**. **c,** The QHE measured at high field region. The second differential of $R_{xx}$ is taken to manifest the SdH oscillations. The Hall conductance $\sigma_{xy}$ is plotted in unit of quantum conductance $\frac{e^2}{h}$ (without degeneracy factor $Zs$). The black arrows are assigned to denote the quantized Hall conductance plateaus. **d,** Temperature dependence of the oscillatory components in the longitudinal MRs. **e,** The extracted effective mass $m^*$ of bulk state and Weyl orbit plotted as a function of $1/\mu_0 H$.